\documentclass{desyproc}

\begin{document}
%------------------------------------
\title{Recent results from the CDMS-II experiment}

%for single authors the superscripts are optional
\author{{\slshape Tobias Bruch$^1$ for the CDMS Collaboration}\\[1ex]
$^1$University of Z\"urich, Winterthurerstr 190, 8047 Z\"urich, Switzerland\\}

% if the proceedings are available online (e.g. at Indico)
% please enter the contribution ID or file_name below for the DOI
%\contribID{32}
\contribID{bruch\_tobias}

% TO THE CONFERENCE EDITORS: 
% please update the following information      
% before sending the template to the authors
% \confID{800}  % if the conference is on Indico uncomment this line
\desyproc{DESY-PROC-2008-02}
\acronym{Patras 2008} % if you want the Acronym in the page footer uncomment this line
\doi  % if there is an online version we will register DOIs

\maketitle

\begin{abstract}
The Cryogenic Dark Matter Search experiment (CDMS) employs low-temperature Ge and Si detectors to detect WIMPs via their elastic scattering interaction with the target nuclei. The current analysis of 397.8\,kg-days Ge exposure resulted in zero observed candidate events, setting an upper limit on the spin-independent WIMP-nucleon cross-section of 6.6\,$\times$\,$10^{-44}$\,cm$^2$ (4.6\,$\times$\,$10^{-44}$\,cm$^2$, when previous CDMS Soudan data is included) at the 90\% confidence level for a WIMP mass of 60\,GeV. To increase the sensitivity, new one inch thick detectors have been developed which will be used in the SuperCDMS phase. SuperCDMS 25kg will be operated at SNOLAB with an expected sensitivity on the spin-independent WIMP-nucleon elastic scattering cross-section of 1\,$\times$\,$10^{-45}$\,cm$^2$.
\end{abstract}
\section{Introduction}
The Cryogenic Dark Matter Search (CDMS) experiment operates 19 Ge (250\,g each) and 11 Si (100\,g each) detectors at the Soudan underground laboratory (MN, USA) to search for non-luminous, non-baryonic Weakly Interacting Massive Particles (WIMPs), that could form the majority of the matter in the universe \cite{Spergel,Jungman} . Each detector is a disk 7.6\,cm in diameter and 1\,cm thick. The detectors are operated at cryogenic temperatures $\sim$ 40 mK to collect the athermal phonons created upon an interaction in the crystal in four independent sensors. In addition, the electron hole pairs created by a recoil are drifted in a field of 3\,V/cm (Ge), 4\,V/cm (Si) towards two concentric electrodes lithographically patterned on one flat side of the crystals \cite{zipdetectors}. In the analysis events from the outer part of the detectors are removed by a fiducial volume cut based on the partitioning of energy between the two concentric charge electrodes.
The simultaneous measurement of the phonon and ionization recoil energy of an interaction in the crystals not only allows an accurate measurement of the recoil energy independent of recoil type (nuclear/electron recoil), but also allows the discrimination between nuclear and electron recoils by the so called ionization yield parameter, which is the ratio of the ionization and phonon energy, providing a rejection factor of $>$\,$10^4$. Nuclear recoils produce fewer charge pairs, and hence less ionization energy than do electron recoils of the same energy. The ionization yield for electron and nuclear recoils is determined from $^{133}$Ba and $^{252}$Cf calibrations respectively, providing the bands shown in Fig.\ref{fig:yieldtiming}.

\section{Backgrounds}
Passive shielding, consisting of lead and polyethylene layers, are used to reduce external gamma and neutron backgrounds, leaving decays of radioactive contamination inside the shielding as the dominant natural background. Monte Carlo simulations with the GEANT4 Toolkit of the radioactive contamination ( isotopes of the $^{238}$U and $^{232}$Th chain as well as $^{60}$Co and $^{40}$K ) of materials inside the experimental apparatus match the observed background spectra very well, revealing no unidentified spectral lines, which would indicate an additional contamination. The whole experimental setup is surrounded by an active scintillator veto to reject events caused by cosmogenic muons or showers. Neutrons induced by radioactive processes or cosmogenic muons interacting near the experimental apparatus, can generate nuclear recoils which cannot be distinguished from possible dark matter interactions on an event-by-event basis. Monte Carlo simulations of both sources give a conservative upper limit of $<$ 0.1 events in the current WIMP-search data from each source.

\begin{figure}[]

\begin{minipage}[c]{0.495\textwidth}
      \centering
\includegraphics[scale=0.38]{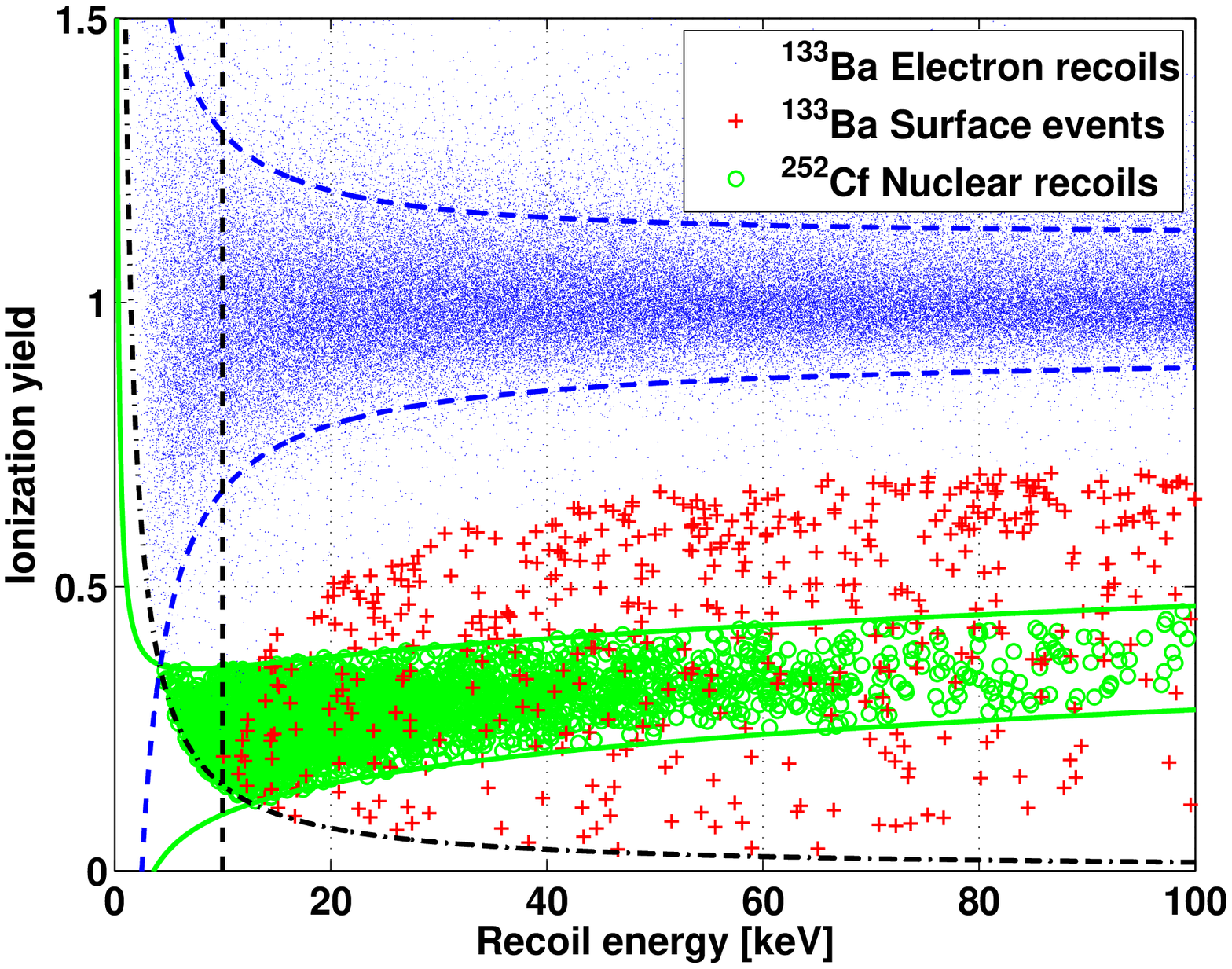}
 \end{minipage}
\begin{minipage}[c]{0.495\textwidth}
      \centering
\includegraphics[scale=0.38]{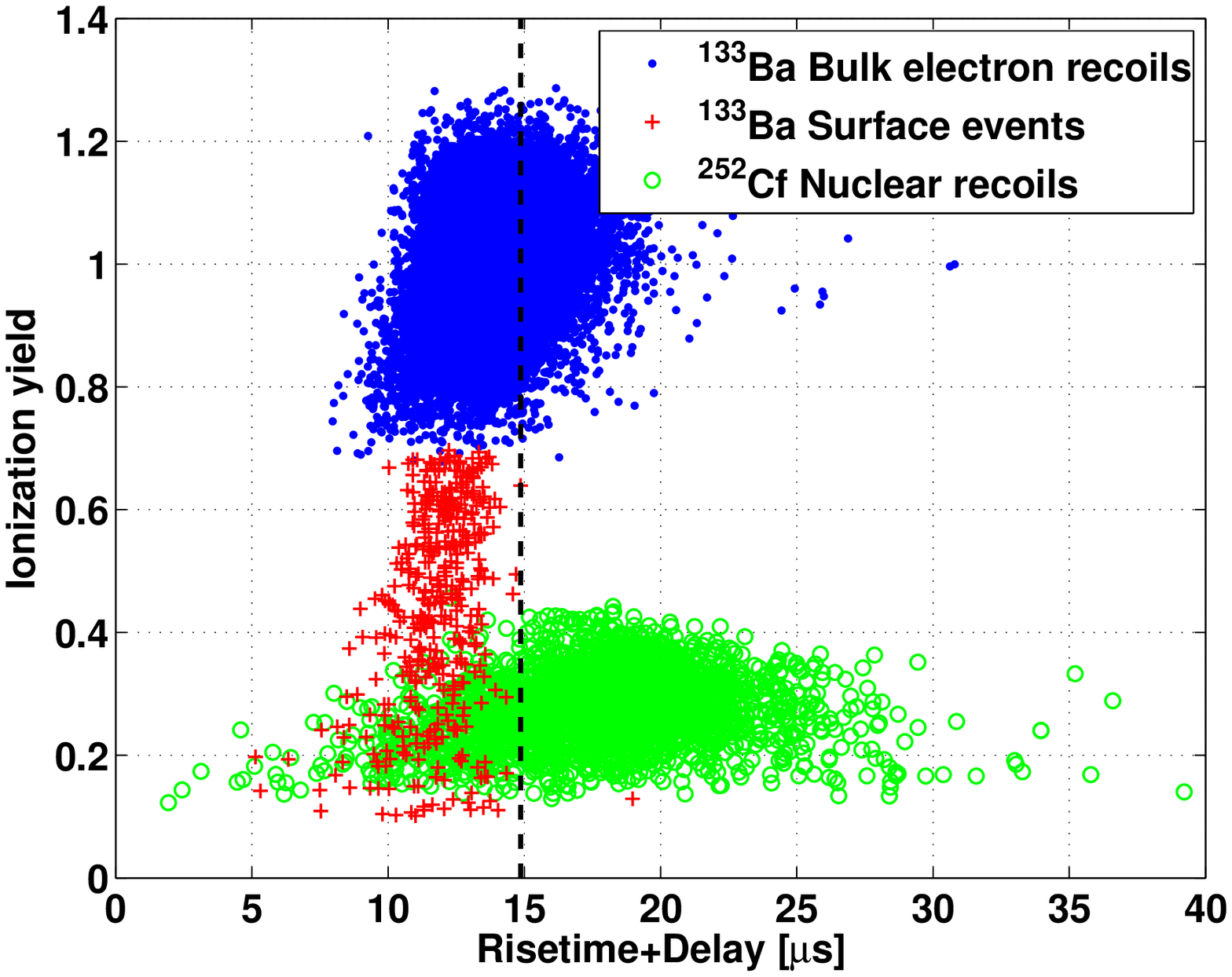}
 \end{minipage}
\caption{\small{Ionization yield as a function of the recoil energy (left panel) and phonon-timing parameter (right panel) for calibration data in a germanium detector. Three different classes of events are identified: $^{133}$Ba gamma-calibration bulk (blue dots) and surface (red crosses) events, as well as nuclear recoils (green circles) from $^{252}$Cf neutron-calibration data. In the left plot the nuclear - (green line) and electron-recoil band (blue/dashed) are shown with the analysis threshold (dashed/vertical) of 10\,keV and the ionization threshold (black/dot dash). The vertical dashed line in the right plot indicates the position of the timing cut, efficiently cutting out surface events.}}\label{fig:yieldtiming}
\end{figure}

Particle interactions may suffer from a suppressed ionization signal if the interactions occur in the first few microns of the crystal surfaces. For events interacting in the first few microns the ionization loss is sufficient to missclassify these as nuclear recoils. These events, referred to as surface events, can be identified as a third population between the electron and nuclear band in Fig.\ref{fig:yieldtiming}. Surface events mainly occur due to radioactive contamination on detector surfaces, or as a result of external gamma ray interactions releasing low-energy electrons from surfaces near the detectors. A correlation analysis between alpha-decay and surface-event rates provides evidence that $^{210}$Pb is a major component of our surface event background. The remaining surface-event rate is compatible with the rate expected from photon induced events \cite{pb210background}. 

To discriminate surface events against nuclear recoil events the timing properties of the phonon pulses are used. The two parameters used to cut out surface events and select nuclear recoils are the delay of the slower phonon signal with respect to the ionization signal and the risetime of the leading phonon pulse (which is the one with the highest amplitude), since surface events have smaller delays and faster risetimes than bulk nuclear-recoils. The cut, based on the sum of risetime and delay provides good surface event rejection, improving the overall rejection of electron recoils to $>$\,$10^{6}$. The cut is designed for each detector independently by using calibration data only as shown in Fig.\ref{fig:yieldtiming}. Only single scatters with a timing parameter value greater than the cut value are considered as WIMP candidates. A single scatter is required to deposit energy in one and only one detector. In the analysis the signal window is constrained by the 2$\sigma$ nuclear recoil band.

\section{Results}
The current analysis used data ( 397.8\,kg-days of germanium exposure ) from two periods (run 123 and 124) between October 2006 and July 2007. Of the 19 Ge detectors, three suffering reduced performance from readout failures and one with relatively poor energy resolution have been left out. The remaining 15 Ge detectors were used for the run 123 analysis. The data taken with eight detectors in run 124 is not considered in this analysis, since they have differences in performance between the two runs. 

\begin{wrapfigure}{l}{0.5\textwidth}
\centering
\includegraphics[scale=0.7]{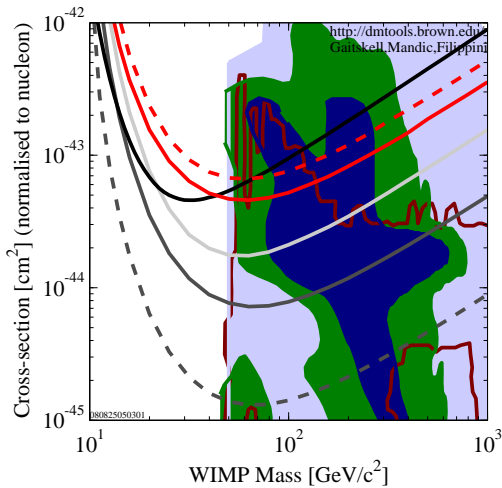}
\caption{\small{Upper limits on the spin independent WIMP-nucleon cross-section from the current analysis (red dashed) and the combined limit by including previous CDMS data (red solid) \cite{r123analysis}. Also shown are the limit from the XENON10 experiment \cite{xenon10} (black solid) and expected sensitivities of the CDMS-II setup until the end of 2008 (light gray); two super towers operated at Soudan (dark gray/solid) and the SuperCDMS 25 kg stage (dark gray/dashed). Filled regions indicate CMSSM models \cite{baltz,ruiz}.}}
\label{fig:exlimits}
\end{wrapfigure}

Surface events present in $^{133}$Ba calibration data or naturally present in WIMP search data, were studied to determine the surface event leakage into the signal region after the timing cut is applied. The estimated surface event leakage, based on the observed numbers of single- and multiple- scatter events within and sourrounding the 2$\sigma$ nuclear recoil region in each detector, is $0.6^{+0.5}_{-0.3} (stat.) ^{+0.3}_{-0.2} (syst.)$ \cite{r123analysis} events. After all analysis cuts were finalized and leakage estimation schemes selected, the single recoil blinded WIMP signal region was unmasked on February 4th, 2008. No event was observed within the signal region.
\begin{wrapfigure}{l}{0.53\textwidth}
\centering
\includegraphics[scale=0.42]{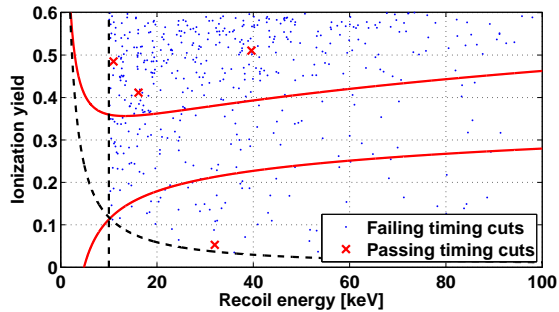}
\caption{\small{Low background events around the 2$\sigma$ nuclear recoil signal region (red band) from all detectors used in this analysis, failing (blue dots) and passing (red crosses) the timing cuts. The four events passing the timing cuts are outside of the nuclear recoil region }}\label{fig:coaddlowback}
\end{wrapfigure}
The WIMP search data of all detectors is shown in Fig.\ref{fig:coaddlowback}. The four events passing the timing cut (red crosses in the figure) are outside of the 2 $\sigma$ nuclear recoil signal region. From this data the 90\% CL upper limit on the spin-independent WIMP-nucleon cross section shown as the red dashed line in Fig. \ref{fig:exlimits} is derived \cite{r123analysis}. The inclusion of a reanalysis of previous CDMS data \cite{rc89analysis} (shown as the red solid line), sets the world's most stringent upper limit on the spin-independent WIMP-nucleon cross section for WIMP masses above 42\,GeV/c$^2$ with a minimum of 4.6\,$\times$\,$10^{-44}$\,cm$^2$ for a WIMP mass of 60\,GeV/c$^2$. 

To further increase the sensitivity the total accumulated exposure has to be increased. This can be achieved by increasing the detector mass and the runtime of the experiment. So far the CDMS-II setup has acquired 1500\,kg-days of Ge raw exposure (including run 123/124) and is expected to accumulate an additional exposure of $\sim$\,500\,kg-days until the end of 2008. For the SuperCDMS setup new 1 inch thick detectors have been developed and tested, providing an increase of a factor 2.54 in mass with respect to the 1\,cm thick detectors used in CDMS-II. The redesign of the phonon readout, which maximizes the active phonon collection area, and new sensor configurations are expected to improve the discrimination between surface events and nuclear-recoils. The first two super towers consisting each of six 1\,inch thick detectors will be installed at the Soudan site by 2009 to demonstrate the improved discrimination capabilities, and show that the operation with a background free signal region can be maintained. At the SuperCDMS 25\,kg stage seven super towers will be installed and operated at SNOLAB. As shown in Fig.\ref{fig:exlimits}, SuperCDMS 25\,kg aims to reach a sensitivity of 1\,$\times$\,$10^{-45}$\,cm$^{2}$ at a WIMP mass of 60\,GeV/c$^2$.

\section{Conclusions}
The CDMS-II experiment has maintained high dark matter discovery potential by limiting expected backgrounds to less than one event in the signal region. The current data sets the world's most stringent upper limit on the spin-independent WIMP-nucleon cross-section for WIMP masses above 42\,GeV/c$^2$ with a minimum of 4.6\,$\times $\,$10^{-44}$\,cm$^2$ for a WIMP mass of 60\,GeV/c$^2$. Ongoing runs aim to accumulate roughly 2000\, kg-days of WIMP search exposure until the end of 2008. By this the CDMS-II experiment is expected to reach a sensitivity of 1\,$\times$\,$10^{-44}$\,cm$^{2}$.
The first two super towers with new 1\,inch thick detectors will be installed at the Soudan site by 2009 demonstrating the improved discrimination capabilities. The next upgrade of the CDMS experiment to SuperCDMS 25\,kg operating seven super towers will be installed at SNOLAB, increasing the sensitivity by one order of magnitude.

% ****************************************************************************
% BIBLIOGRAPHY AREA
% ****************************************************************************

\begin{footnotesize}
% IF YOU DO NOT USE BIBTEX, USE THE FOLLOWING SAMPLE SCHEME FOR THE REFERENCES
% ----------------------------------------------------------------------------

% ----------------------------------------------------------------------------

% IF YOU USE BIBTEX,
% - DELETE THE TEXT BETWEEN THE TWO ABOVE DASHED LINES
% - UNCOMMENT THE NEXT TWO LINES AND REPLACE 'Name_Of_Your_BibFile'

%\bibliographystyle{unsrt}
%\bibliography{Name_Of_Your_BibFile}
% example of Name_Of_Your_BibFile.bib
% @Article{Turcato:2006ch,
%      author    = "Turcato, M.",
%  collaboration = "ZEUS and H1",
%      title     = "Lepton flavour violation and charmonium physics at HERA",
%      journal   = "Nucl. Phys. Proc. Suppl.",
%      volume    = "162",
%      year      = "2006", 
%      pages     = "283-287",
%      SLACcitation  = "%%CITATION = NUPHZ,162,283;%%"
% }
% 
% @Unpublished{Gogitidze:2007du,
%      author    = "Gogitidze, N.",
%  collaboration = "H1", 
%      title     = "Prompt photons and particle momentum distributions at
%                   HERA", 
%      year      = "2007",
%      note    = "hep-ex/0701033",
%      SLACcitation  = "%%CITATION = HEP-EX 0701033;%%"
% }

\end{footnotesize}

% ****************************************************************************
% END OF BIBLIOGRAPHY AREA
% ****************************************************************************

\end{document}